\def\year{2019}\relax
\documentclass[letterpaper]{article} 
\usepackage{aaai19}  
\usepackage{times}  
\usepackage{helvet}  
\usepackage{courier}  
\usepackage{url}  
\usepackage{graphicx}  
\usepackage{amsmath}
\usepackage{amsthm}
\theoremstyle{remark}
\newtheorem{myDef}{Definition}
\newtheorem{myExp}{Example}
\usepackage{amsfonts}
\usepackage{multirow}
\usepackage{makecell}
\usepackage{booktabs}
\usepackage{mathtools}
\usepackage[ruled]{algorithm2e} 
\usepackage{graphicx}

\begin{document}
	\title{Relation Structure-Aware Heterogeneous Information Network Embedding}
	\author{Yuanfu Lu\textsuperscript{1}, Chuan Shi\textsuperscript{1}\thanks{Corresponding author: Chuan Shi (shichuan@bupt.edu.cn).}, Linmei Hu\textsuperscript{1}, Zhiyuan Liu\textsuperscript{2} \\
		\textsuperscript{1}Beijing University of Posts and Telecommunications, Beijing, China \\
		\textsuperscript{2}Tsinghua University, Beijing, China \\
		\{luyuanfu,  shichuan, hulinmei\}@bupt.edu.cn, liuzy@tsinghua.edu.cn\\
	}

\maketitle
\setcounter{secnumdepth}{2}

\begin{abstract}
	Heterogeneous information network (HIN) embedding aims to embed multiple types of nodes into a low-dimensional space. Although most existing HIN embedding methods consider heterogeneous relations in HINs, they usually employ one single model for all relations without distinction, which inevitably restricts the capability of network embedding. 
	In this paper, we take the structural characteristics of heterogeneous relations into consideration and propose a novel Relation structure-aware Heterogeneous Information Network Embedding model (RHINE). 
	By exploring the real-world networks with thorough mathematical analysis, we present two structure-related measures which can consistently distinguish heterogeneous relations into two categories: Affiliation Relations (ARs) and Interaction Relations (IRs). 
	To respect the distinctive characteristics of relations, in our RHINE, we propose different models specifically tailored to handle ARs and IRs, which can better capture the structures and semantics of the networks. At last, we combine and optimize these models in a unified and elegant manner.
	Extensive experiments on three real-world datasets demonstrate that our model significantly outperforms the state-of-the-art methods in various tasks, including node clustering, link prediction, and node classification.
\end{abstract}

\section{Introduction}
\begin{figure}[!tpb]
	\centering
	\includegraphics[width=1\linewidth]{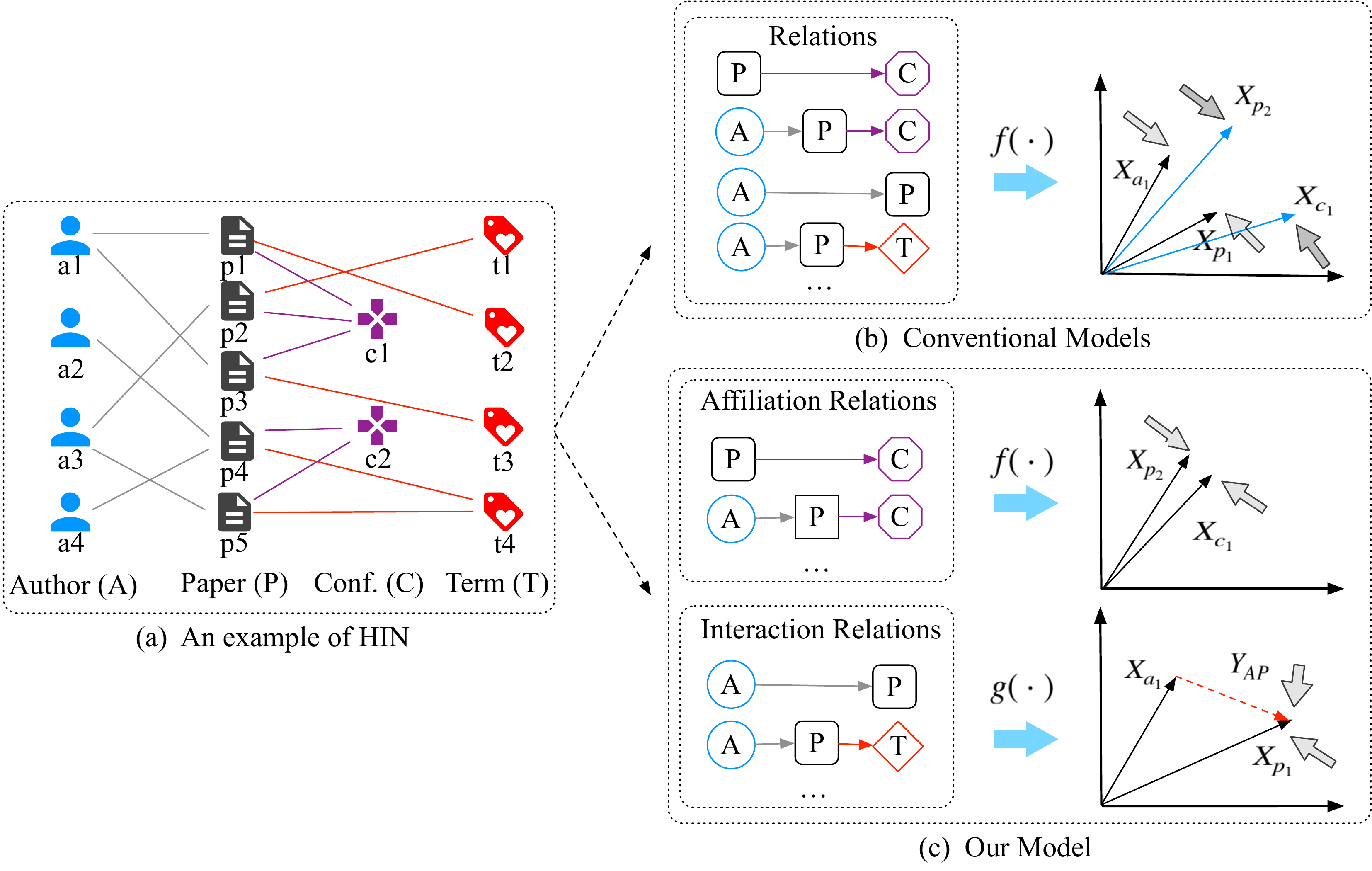}
	\caption{The illustration of an HIN and the comparison between conventional methods and our method  (non-differentiated relations v.s. differentiated relations).}
	\label{fig:one}
\end{figure}
Network embedding has shed a light on the analysis of networks as it is effective to learn the latent features that encode the properties of a network \cite{cui2018survey,cai2018comprehensive}. 
Although the state-of-the-arts \cite{perozzi2014deepwalk,grover2016node2vec,tang2015line,wang2016structural} have achieved promising performance in many data mining tasks, most of them focus on homogeneous networks, which only contain one single type of nodes and edges. 
In reality, many networks are usually with multiple types of nodes and edges, widely known as heterogeneous information networks (HINs) \cite{sun2011pathsim,shi2017survey}.
Taking the DBLP network for example, as shown in Figure~\ref{fig:one}(a), it contains four types of nodes: Author (A), Paper (P), Conference (C) and Term (T), and multiple types of relations: writing/written relations, and publish/published relations, etc. In addition, there are composite relations represented by meta-paths \cite{sun2011pathsim}  such as \textit{APA} (co-author relation) and \textit{APC} (authors write papers published in conferences), which are widely used to exploit rich semantics in HINs. 
Thus, compared to homogeneous networks, HINs fuse more information and contain richer semantics. Directly applying traditional homogeneous models to embed HINs will inevitably lead to reduced performance in downstream tasks.

To model the heterogeneity of networks, several attempts have been done on HIN embedding. 
For example, some models employ meta-path based random walk to generate node sequences for optimizing the similarity between nodes  \cite{shang2016meta,dong2017metapath2vec,fu2017hin2vec}. 
Some methods decompose the HIN into simple networks and then optimize the proximity between nodes in each sub-network \cite{tang2015pte,xu2017embedding,shi2018heterogeneous}. 
There are also some neural network based methods that learn non-linear mapping functions for HIN embedding \cite{chang2015heterogeneous,wang2018shine,han2018aspect}. 
Although these methods consider the heterogeneity of networks, they usually have an assumption that one single model can handle all relations and nodes, through keeping the representations of two nodes close to each other, as illustrated in Figure \ref{fig:one}(b). 

However, various relations in an HIN have significantly different structural characteristics, which should be handled with different models. Let's see a toy example in Figure~\ref{fig:one}(a). The relations in the network include atomic relations (e.g., \textit{AP} and \textit{PC}) and composite relations (e.g., \textit{APA} and \textit{APC}). 
Intuitively, \textit{AP} relation and \textit{PC} relation reveal rather different characteristics in structure. That is, some authors write some papers in the \textit{AP} relation, which shows a peer-to-peer structure. While that many papers are published in one conference in the \textit{PC} relation reveals the structure of one-centered-by-another. 
Similarly, \textit{APA} and \textit{APC} indicate peer-to-peer and one-centered-by-another structures respectively.  The intuitive examples clearly illustrate that relations in an HIN indeed have different structural characteristics. 

It is non-trivial to consider different structural characteristics of relations for HIN embedding, due to the following challenges: 
(1) How to distinguish the structural characteristics of relations in an HIN? 
Various relations (atomic relations or meta-paths) with different structures are involved in an HIN. Quantitative and explainable criteria are desired to explore the structural characteristics of relations and distinguish them. 
(2) How to capture the distinctive structural characteristics of different categories of relations? Since the various relations have different structures, modeling them with one single model may lead to some loss of information. We need to specifically design appropriate models which are able to capture their distinctive characteristics. 
(3) The different  models for the differentiated relations should be easily and smoothly combined to ensure simple optimization in a unified manner. 

In this paper, we present a novel model for HIN embedding, named \textbf{R}elation structure-aware \textbf{HIN} \textbf{E}mbedding (\textbf{RHINE}). 
In specific, we first explore the structural characteristics of relations in HINs with thorough mathematical analysis, and present two structure-related measures which can consistently distinguish the various relations into two categories: Affiliation Relations (ARs) with one-centered-by-another structures and Interaction Relations (IRs) with peer-to-peer structures. 
In order to capture the distinctive structural characteristics of the relations, we then propose two specifically designed models.
For ARs where the nodes share similar properties \cite{yang2012community}, we calculate Euclidean distance as the proximity between nodes, so as to make the nodes directly close in the low-dimensional space. On the other hand,  for IRs which bridge two compatible nodes, we model them as translations between the nodes.
Since the two models are consistent in terms of  mathematical form, they can be optimized in a unified and elegant way.

It is worthwhile to highlight our contributions as follows:
\begin{itemize}
	\item 
	To the best of our knowledge, we make the first attempt to explore the different structural characteristics of relations in HINs and present two structure-related criteria which can consistently distinguish heterogeneous relations into ARs and IRs.
	\item 
	We propose a novel relation structure-aware HIN embedding model (RHINE), which fully respects the distinctive structural characteristics of ARs and IRs by exploiting appropriate models and combining them in a unified and elegant manner.
	\item 
	We conduct comprehensive experiments to evaluate the performance of our model. Experimental results demonstrate that our model significantly outperforms state-of-the-art network embedding models in various tasks.
\end{itemize}
\begin{table*} [htbp] 
	\caption{Statistics of the three datasets. $ t_u$ denotes the type of node $u$, $ \langle u, r, v \rangle $ is a node-relation triple.}
	\label{tab:one}
	\centering
	\resizebox{0.9\textwidth}{30mm}{
		\begin{tabular}{ccc|cccc|cc|c}
			\toprule
			\multirow{2}*{Datasets}&\multirow{2}*{Nodes}&Number of&Relations&Number of&Avg. Degree & Avg. Degree & \multicolumn{2}{c|}{Measures} & Relation \\
			~ & ~ & Nodes & ($ t_u \sim  t_v $) & Relations & of $t_u $ & of $t_v $ & $ D(r) $ & $ S(r) $ & Category \\
			
			\midrule
			DBLP 
			& \makecell[c]{Term (T) \\ Paper (P) \\ Author (A)  \\ Conference (C)} 
			& \makecell[c]{ 8,811 \\ 14,376 \\ 14,475 \\ 20 } 
			& \makecell[c]{PC \\ APC \\ AP \\ PT \\ APT } 
			& \makecell[c]{14,376 \\ 24,495 \\ 41,794 \\ 88,683 \\ 260,605 } 
			& \makecell[c]{1.0 \\ 2.9 \\ 2.8 \\ 6.2 \\ 18.0 } 
			& \makecell[c]{718.8 \\ 2089.7 \\ 2.9 \\ 10.7 \\ 29.6 } 
			& \makecell[c]{718.8\\ 720.6 \\ 1.0 \\ 1.7 \\ 1.6 } 
			& \makecell[c]{0.05\\ 0.085 \\ 0.0002 \\ 0.0007 \\ 0.002 } 
			& \makecell[c]{AR\\ AR\\ IR \\ IR \\ IR } \\
			\midrule
			
			Yelp 
			& \makecell[c]{User (U) \\ Service (S) \\ Business (B) \\ Star Level (L)  \\ Reservation (R)} 
			& \makecell[c]{ 1,286 \\ 2 \\ 2,614 \\ 9 \\ 2  } 
			& \makecell[c]{BR \\ BS \\ BL \\ UB \\ BUB } 
			& \makecell[c]{2,614 \\ 2,614 \\ 2,614 \\ 30,838 \\ 528,332 } 
			& \makecell[c]{1.0 \\ 1.0 \\ 1.0 \\ 23.9 \\ 405.3  } 
			& \makecell[c]{1307.0 \\ 1307.0 \\ 290.4 \\ 11.8 \\ 405.3 } 
			& \makecell[c]{1307.0 \\ 1307.0 \\ 290.4  \\ 2.0 \\ 1.0 } 
			& \makecell[c]{0.5\\ 0.5 \\ 0.1 \\ 0.009 \\  0.07 } 
			& \makecell[c]{AR\\ AR\\ AR\\ IR \\ IR } \\
			\midrule
			AMiner 
			& \makecell[c]{Paper (P) \\ Author (A) \\ Reference (R) \\ Conference (C)} 
			& \makecell[c]{ 127,623 \\ 164,472 \\ 147,251 \\ 101 } 
			& \makecell[c]{PC \\ APC \\ AP \\ PR \\ APR } 
			& \makecell[c]{127,623 \\ 232,659 \\ 355,072 \\ 392,519 \\ 1,084,287 } 
			& \makecell[c]{1.0 \\ 2.2 \\ 2.2 \\ 3.1 \\ 7.1 } 
			& \makecell[c]{1263.6 \\ 3515.6 \\ 2.8 \\ 2.7 \\ 7.9 } 
			& \makecell[c]{1264.6\\ 1598.0 \\ 1.3 \\ 1.1 \\ 1.1 } 
			& \makecell[c]{0.01\\ 0.01 \\ 0.00002 \\ 0.00002 \\ 0.00004 } 
			& \makecell[c]{AR\\ AR\\ IR \\ IR \\ IR } \\
			
			\bottomrule
		\end{tabular}
	}
\end{table*}
\section{Related Work}
Recently, network embedding has attracted considerable attention. 
Inspired by word2vec \cite{mikolov2013distributed}, random walk based methods \cite{perozzi2014deepwalk,grover2016node2vec} have been proposed to learn representations of networks by the skip-gram model. 
After that, several models are designed to better preserve network properties \cite{tang2015line,ou2016asymmetric,ribeiro2017struc2vec}. 
Besides, there are some deep neural network based models for network embedding \cite{wang2016structural,cao2016deep}.
However, all the aforementioned methods focus only on learning the representations of homogeneous networks.

Different from homogeneous networks, HINs consist of multiple types of nodes and edges. Several attempts have been done on HIN embedding and achieved promising performance in various tasks \cite{tang2015pte,shang2016meta,fu2017hin2vec,wang2018shine,shi2018heterogeneous}. 
PTE \cite{tang2015pte} decomposes an HIN to a set of bipartite networks and then performs network embedding individually. 
ESim \cite{shang2016meta} utilizes user-defined meta-paths as guidance to learn node embeddings. 
Metapath2vec \cite{dong2017metapath2vec} combines meta-path based random walks and skip-gram model for HIN embedding. 
HIN2Vec \cite{fu2017hin2vec} learns the embeddings of HINs by conducting multiple prediction training tasks jointly. 
HERec \cite{shi2018heterogeneous} filters the node sequences with type constraints and thus captures the semantics of HINs.

All the above-mentioned models deal with all relations in HINs with one single model, neglecting differentiated structures of relations. 
In this paper, we explore and distinguish the structural characteristics of relations with quantitative analysis. For relations with distinct structural characteristics, we propose to handle them with specifically designed models.
\section{Preliminaries}
In this section, we introduce some basic concepts and formalize the problem of HIN embedding.
\begin{myDef}\textit{Heterogeneous Information Network (HIN).}
	An HIN is defined as a graph $G = (V, E, T, \phi, \varphi)$, in which $V$ and $E$ are the sets of nodes and edges, respectively. Each node $v$ and edge $e$ are associated with their type mapping functions $\phi: V\to {T_V}$ and $\varphi: E\to {T_E} $, respectively. $ T_V $ and $ T_E $ denote the sets of node and edge types, where $ |T_V|+ |T_E|>2$, and $ T = T_{V} \cup{T_E} $.
\end{myDef}

\begin{myDef}\textit{Meta-path}. 
	A meta-path $ m\in{M} $ is defined as a sequence of node types $ t_{v_{i}} $ or edge types $ t_{e_{j}} $ in the form of $ t_{v_1}\stackrel{t_{e_1}}{\longrightarrow}t_{v_2} ... \stackrel{t_{e_l}}{\longrightarrow}t_{v_{l+1}} $ (abbreviated as $ t_{v_1}t_{v_2}...t_{v_l+1} $ ), which describes a composite relation between $ v_1 $ and $ v_{l+1} $.
	
\end{myDef}

\begin{myDef}\textit{Node-Relation Triple}. 
	In an HIN $ G $, relations $ R $ include atomic relations (e.g., links) and composite relations (e.g., meta-paths). A node-relation triple $ \langle u, r, v \rangle \in{P}$, describes that two nodes $ u $ and $ v $ are connected by a relation $ r\in{R} $. Here $ P $ represents the set of all node-relation triples. 
	
\end{myDef}

\begin{myExp}
	For example, as shown in Figure \ref{fig:one}(a), $ \langle a_2, APC,c_2 \rangle $ is a node-relation triple, meaning that $ a_1 $ writes a paper published in $ c_2 $.
\end{myExp}

\begin{myDef}\textit{Heterogeneous Information Network Embedding}. 
	Given an HIN $ G=(V$, $E$, $T$, $\phi$, $\varphi)$ , the goal of HIN embedding is to develop a mapping function $ f: V \to {\mathbb{R}^d} $ that projects each node $ v\in{V} $ to a low-dimensional vector in $ \mathbb{R}^d $, where $ d \ll |V| $. 
\end{myDef}

\section{Structural Characteristics of Relations}
In this section, we first describe three real-world HINs and analyze the structural characteristics of relations in HINs. Then we present two structure-related measures which can consistently distinguish various relations quantitatively. 

\subsection{Dataset Description}\label{dataset}
Before analyzing the structural characteristics of relations, we first briefly introduce three datasets used in this paper, including DBLP\footnote[1]{https://dblp.uni-trier.de}, Yelp\footnote[2]{https://www.yelp.com/dataset/} and AMiner\footnote[3]{https://www.aminer.cn/citation}\cite{tang2008arnetminer}. The detailed statistics of these datasets are illustrated in Table ~\ref{tab:one}. 

DBLP is an academic network, which contains four types of nodes: author (A), paper (P), conference (C) and term (T). We extract node-relation triples based on the set of relations \{\textit{AP, PC, PT, APC, APT}\}. Yelp is a social network, which contains five types of nodes: user (U), business (B), reservation (R), service (S) and star level (L). We consider the relations \{\textit{BR, BS, BL, UB, BUB}\}. AMiner is also an academic network, which contains four types of nodes, including author (A), paper (P), conference (C) and reference (R). We consider the relations \{\textit{AP, PC, PR, APC, APR}\}. Notice that we can actually analyze all the relations based on meta-paths. However, not all meta-paths have a positive effect on embeddings \cite{sun2013pathselclus}. Hence, following previous works \cite{shang2016meta,dong2017metapath2vec}, we choose the important and meaningful meta-paths.

\subsection{Affiliation Relations and Interaction Relations}
In order to explore the structural characteristics of relations, we present mathematical analysis on the above datasets.

Since the degree of nodes can well reflect the structures of networks \cite{wasserman1994social}, we define a degree-based measure $ D(r)$ to explore the distinction of various relations in an HIN. Specifically, we compare the average degrees of two types of nodes connected with the relation $ r $, via dividing the larger one by the smaller one ($ D(r) \ge 1 $). 
Formally, given a relation $ r $ with nodes $ u $ and $ v $ (i.e., node relation triple $ \langle u, r, v \rangle $),  $ t_{u} $ and $ t_{v} $ are the  node types of $ u $ and $ v $, we define $ D(r) $ as follows:
\begin{equation} 
\fontsize{9pt}{10pt}
D(r)= \frac{\max{[\bar{d}_{t_u},\bar{d}_{t_v}]}}{\min{[\bar{d}_{t_u},\bar{d}_{t_v}]}},
\label{equation:one}
\end{equation}
where $ \bar{d}_{t_u} $ and $ \bar{d}_{t_v} $ are the average degrees of nodes of the types $ t_u $ and $ t_v $ respectively. 

A large value of $ D(r)$ indicates quite inequivalent structural roles of two types of nodes connected via the relation $ r $ (one-centered-by-another), while a small value of $ D(r) $ means compatible structural roles (peer-to-peer). In other words, relations with a large value of $ D(r) $ show much stronger affiliation relationships. Nodes connected via such relations share much more similar properties \cite{faust1997centrality}. While relations with a small value of $ D(r) $ implicate much stronger interaction relationships. 
Therefore, we call the two categories of relations as \textit{Affiliation Relations} (ARs) and \textit{Interaction Relations} (IRs), respectively.

In order to better understand the structural difference between various relations, we take the DBLP network as an example. As shown in Table \ref{tab:one}, for the relation \textit{PC} with $ D(PC)=718.8$, the average degree of nodes with type P is 1.0 while that of nodes with type C is 718.8. It shows that papers and conferences are structurally inequivalent. Papers are centered by conferences. While $ D(AP) =1.1 $ indicates that authors and papers are compatible and peer-to-peer in structure. This is consistent with our common sense. Semantically, the relation \textit{PC} means that \textit{`papers are published in conferences'}, indicating an affiliation relationship. Differently, \textit{AP} means that \textit{`authors write papers'}, which explicitly describes an interaction relationship. 

In fact, we can also define some other measures to capture the structural difference. For example, we compare the relations in terms of sparsity, which can be defined as:
\begin{equation}
S(r)=\frac{N_r}{N_{t_u}\times{N_{t_v}}},
\end{equation} 
where $ N_r $ represents the number of relation instances following $ r $. $ N_{t_u} $ and $ N_{t_v} $ mean the number of nodes with type $ t_u $ and $ t_v $, respectively. The measure can also consistently distinguish the relations into two categories: ARs and IRs. The detailed statistics of all the relations in the three HINs are shown in Table ~\ref{tab:one}. 

Evidently, Affiliation Relations and Interaction Relations exhibit rather distinct characteristics: (1) ARs indicate one-centered-by-another structures, where the average degrees of the types of end nodes are extremely different. They imply an affiliation relationship  between nodes. (2) IRs describe peer-to-peer structures, where the average degrees of the types of end nodes are compatible. They suggest an interaction relationship between nodes. 

\section{Relation Structure-Aware HIN Embedding}
In this section, we present a novel Relation structure-aware HIN Embedding model (RHINE), which individually handles two categories of relations  (ARs and IRs) with different models in order to preserve their distinct structural characteristics, as illustrated in Figure 1(c).

\subsection{Basic Idea}
Through our exploration with thorough mathematical analysis, we find that the heterogeneous relations can be typically divided into ARs and IRs with different structural characteristics. In order to respect their distinct characteristics, we need to specifically design different while appropriate models for the different categories of relations.

For ARs, we propose to take Euclidean distance as a metric to measure the proximity of the connected nodes in the low-dimensional space. There are two motivations behind this: 
(1) First of all, ARs show affiliation structures between nodes, which indicate that nodes connected via such relations share similar properties. \cite{faust1997centrality,yang2012community}. Hence, nodes connected via ARs could be directly close to each other in the vector space, which is also consistent with the optimization of Euclidean distance \cite{danielsson1980euclidean}. 
(2) Additionally, one goal of HIN embedding is to preserve the high-order proximity. Euclidean distance can ensure that both first-order and second-order proximities are preserved as it meets the condition of the triangle inequality \cite{hsieh2017collaborative}. 

Different from ARs, IRs indicate strong interaction relationships between compatible nodes, which themselves contain important structural information of two nodes. 
Thus, we propose to explicitly model an IR as a translation between nodes in the low-dimensional vector space. Additionally, the translation based distance is consistent with the Euclidean distance in the mathematical form \cite{bordes2013translating}. Therefore, they can be smoothly combined in a unified and elegant manner.

\subsection{Different Models for ARs and IRs}
In this subsection, we introduce two different models exploited in RHINE for ARs and IRs, respectively.

\subsubsection{Euclidean Distance for Affiliation Relations}
Nodes connected via ARs share similar properties \cite{faust1997centrality}, therefore nodes could be directly close to each other in the vector space.
We take the Euclidean distance as the proximity measure of two nodes connected by an AR. 

Formally, given an affiliation node-relation triple $ \langle p, s, q \rangle \in {P_{AR}} $ where $ s\in R_{AR} $ is the relation between $ p $
and $ q $ with weight $ w_{pq} $, the distance between $ p $ and $ q $ in the latent vector space is calculated as follows:
\begin{equation}
f(p, q) = w_{pq} ||\textbf{X}_p-\textbf{X}_q||^2_2,
\end{equation}
in which $ \textbf{X}_p \in{\mathbb{R}^{d}}  $ and $ \textbf{X}_q  \in{\mathbb{R}^{d}}$ are the embedding vectors of $ p $ and $ q $, respectively. 
As $ f(p, q) $ quantifies the distance between $ p $ and $ q $ in the low-dimensional vector space, we aim to minimize $ f(p, q) $ to ensure that nodes connected by an AR should be close to each other. Hence, we define the margin-based loss \cite{bordes2013translating} function as follows:
\begin{equation}
\begin{split}
L_{EuAR} & = { \sum\limits_{s\in{R_{AR}}}\sum\limits_{ \langle p, s, q \rangle \in{P_{AR}}}} \\
& {\sum\limits_{ \langle p',s,q' \rangle \in{P_{AR}'}}{\max [0,\gamma+f(p, q)-f(p', q')]}},
\end{split}
\end{equation}
where $ \gamma>0 $ is a margin hyperparameter. $ P_{AR} $ is the set of positive affiliation node-relation triples, while $ P_{AR}' $ is the set of negative affiliation node-relation triples.

\subsubsection{Translation-based Distance for Interaction Relations}
Interaction Relations demonstrate strong interactions between nodes with compatible structural roles. Thus, different from ARs, we explicitly model IRs as translations between nodes. 

Formally, given an interaction node-relation triple $ \langle u, r, v \rangle $ where $ r\in R_{IR} $ with weight $ w_{uv} $, we define the score function as:
\begin{equation}
g (u, v) = w_{uv} ||\textbf{X}_u + \textbf{Y}_r -\textbf{X}_v||,
\end{equation}
where $ \textbf{X}_u $ and $ \textbf{X}_v $ are the node embeddings of $ u $ and $ v $ respectively, and $ \textbf{Y}_r $ is the embedding of the relation $ r $. Intuitively, this score function penalizes deviation of $ (\textbf{X}_u+\textbf{Y}_r) $ from the vector $ \textbf{X}_v $.

For each interaction node-relation triple $ \langle u, r, v \rangle \in{P_{IR}} $, we define the margin-based loss function as follows:
\begin{equation}
\begin{split}
L_{TrIR} & = {\sum\limits_{r\in{R_{IR}}} {\sum\limits_{ \langle u, r, v \rangle \in{P_{IR}}}}} \\
& { \sum\limits_{ \langle u', r, v' \rangle \in{P_{IR}'}}{\max [0,\gamma+g(u, v)-g(u', v')]}}
\end{split}
\end{equation}
where $ P_{IR} $ is the set of positive interaction node-relation triples, while $ P_{IR}' $ is the set of negative interaction node-relation triples.

\subsection{A Unified Model for HIN Embedding}
Finally, we smoothly combine the two models for different categories of relations by minimizing the following loss function:
\begin{equation}
\fontsize{9pt}{10pt}
L = L_{EuAR}+L_{TrIR} 
\end{equation}
\begin{equation}
\fontsize{8pt}{10pt}
\begin{split}
& =  { \sum\limits_{s\in{R_{AR}}}\sum\limits_{ \langle p, s, q \rangle \in{P_{AR}}} \sum\limits_{ \langle p',s,q' \rangle \in{P_{AR}'}}{\max [0,\gamma+f(p, q)-f(p', q')]}} \notag \\
& + { \sum\limits_{r\in{R_{IR}}}\sum\limits_{ \langle u, r, v \rangle \in{P_{IR}}}\sum\limits_{ \langle u', r, v' \rangle \in{P_{IR}'}}{\max [0,\gamma+g(u, v)-g(u', v')]}}
\end{split}
\label{equation:loss}
\end{equation}

\subsubsection{Sampling Strategy}
As shown in Table~\ref{tab:one}, the distributions of ARs and IRs are quite unbalanced. What's more, the proportion of relations are unbalanced within ARs and IRs. 
Traditional edge sampling may suffer from under-sampling for relations with a small amount or over-sampling for relations with a large amount. To address the problems, we draw positive samples according to their probability distributions. 
As for negative samples, we follow previous work \cite{bordes2013translating} to construct a set of negative node-relation triples $P'_{(u,r,v)} =  \{(u', r, v) | u'\in V\} \cup \{(u, r, v') | v'\in V \} $ for the positive node-relation triple $ (u,r,v) $, where either the head or tail is replaced by a random node, but not both at the same time.

\section{Experiments}
In this section, we conduct extensive experiments to demonstrate the effectiveness of our model RHINE. 

\subsection{Datasets}
As described in Subsection \ref{dataset}, we conduct experiments on three datasets, including DBLP, Yelp and AMiner. The statistics of them are summarized in Table~\ref{tab:one}.

\subsection{Baseline Methods}
We compare our proposed model RHINE with six state-of-the-art network embedding methods. 
\begin{itemize}
	\item 
	\textbf{DeepWalk} \cite{perozzi2014deepwalk} performs a random walk on networks and then learns low-dimensional node vectors via the skip-gram model. 
	\item 
	\textbf{LINE} \cite{tang2015line} considers first-order and second-order proximities in networks. We denote the model  that only uses first-order or second-order proximity as LINE-1st or LINE-2nd, respectively.
	\item 
	\textbf{PTE} \cite{tang2015pte} decomposes an HIN to a set of bipartite networks and then learns the low-dimensional representation of the network.
	\item 
	\textbf{ESim} \cite{shang2016meta} takes a given set of meta-paths as input to learn a low-dimensional vector space. For a fair comparison, we use the same meta-paths with equal weights in Esim and our model RHINE.
	\item 
	\textbf{HIN2Vec} \cite{fu2017hin2vec} learns the latent vectors of nodes and meta-paths in an HIN by conducting multiple prediction training tasks jointly.
	\item 
	\textbf{Metapath2vec} \cite{dong2017metapath2vec} leverages meta-path based random walks and skip-gram model to perform node embedding. We leverage the meta-paths APCPA, UBSBU and APCPA in DBLP, Yelp and AMiner respectively, which perform best in the evaluations.
\end{itemize}

\subsubsection{Parameter Settings}
For a fair comparison, we set the embedding dimension $ d=100 $ and the size of negative samples $ k=3 $ for all models. For DeepWalk, HIN2Vec and metapath2vec, we set the number of walks per node $ w=10 $, the walk length $ l=100 $ and the window size $ \tau = 5 $. For our model RHINE, the margin $ \gamma $ is set to 1.

\subsection{Node Clustering}
\begin{table}
	\centering
	\caption{Performance Evaluation of Node Clustering.}
	\label{tab:node-clustering}
	\begin{tabular}{cccc}
		\toprule
		Methods					& DBLP				& Yelp									& AMiner \\
		\midrule
		DeepWalk				& 0.3884					& 0.3043					& 0.5427 \\
		LINE-1st				& 0.2775						& 0.3103					& 0.3736 \\
		LINE-2nd			& 0.4675			 			& 0.3593					& 0.3862 \\
		PTE					& 0.3101				 			& 0.3527					& 0.4089 \\
		ESim				& 0.3449							& 0.2214					& 0.3409 \\
		HIN2Vec			& 0.4256						 	& 0.3657					& 0.3948 \\
		metapath2vec & 0.6065							& 0.3507					& 0.5586 \\
		RHINE					& \textbf{0.7204}				& \textbf{0.3882}		& \textbf{0.6024} \\
		\bottomrule
	\end{tabular}

\end{table}

\begin{table*}
	\centering
	\caption{Performance Evaluation of Link Prediction.}
	\label{tab:lp}
	\begin{tabular}{ccccccccccc}
		\toprule
		Methods & \multicolumn{2}{c}{DBLP (A-A)}  & \multicolumn{2}{c}{DBLP (A-C)} & \multicolumn{2}{c}{Yelp (U-B)} & \multicolumn{2}{c}{AMiner (A-A)} & \multicolumn{2}{c}{AMiner (A-C)}\\
		\midrule
		~ 	& AUC & F1 & AUC & F1 & AUC	& F1 & AUC & F1 & AUC & F1\\
		\midrule
		DeepWalk & 0.9131 & 0.8246 & 0.7634 & 0.7047 & 0.8476 & 0.6397 & 0.9122 & 0.8471 & 0.7701 & 0.7112 \\
		LINE-1st & 0.8264 & 0.7233 & 0.5335 & 0.6436 & 0.5084 & 0.4379 & 0.6665 & 0.6274 & 0.7574 & 0.6983 \\
		LINE-2nd & 0.7448 & 0.6741 & 0.8340 & 0.7396 & 0.7509 & 0.6809 & 0.5808 & 0.4682 & 0.7899 & 0.7177 \\
		PTE & 0.8853 & 0.8331 & 0.8843 & 0.7720 & 0.8061 & 0.7043 & 0.8119 & 0.7319 & 0.8442 & 0.7587 \\
		ESim & 0.9077 & 0.8129 & 0.7736 & 0.6795 & 0.6160 & 0.4051 & 0.8970 & 0.8245 & 0.8089 & 0.7392 \\
		HIN2Vec & 0.9160 & 0.8475 & 0.8966 & 0.7892 & 0.8653 & 0.7709 & 0.9141 & 0.8566 & 0.8099 & 0.7282 \\
		metapath2vec & 0.9153 & 0.8431 & 0.8987 & 0.8012 & 0.7818 & 0.5391 & 0.9111 & 0.8530 & 0.8902 & 0.8125 \\
		RHINE & \textbf{0.9315} & \textbf{0.8664} & \textbf{0.9148} & \textbf{0.8478} & \textbf{0.8762} & \textbf{0.7912} & \textbf{0.9316} & \textbf{0.8664} & \textbf{0.9173} & \textbf{0.8262} \\
		\bottomrule
	\end{tabular}
\end{table*}

\begin{table*}
	\centering
	\caption{Performance Evaluation of Multi-class Classification.}
	\label{tab:classification}
	\begin{tabular}{ccccccc}
		\toprule
		Methods & \multicolumn{2}{c}{DBLP}  & \multicolumn{2}{c}{Yelp} & \multicolumn{2}{c}{AMiner} \\
		\midrule
		~ 						& Macro-F1 					& Micro-F1 						& Macro-F1 					& Micro-F1 					& Macro-F1 							& Micro-F1 \\
		\midrule
		DeepWalk 			& 0.7475 					& 0.7500 				& 0.6723 					& 0.7012 						& 0.9386 				& 0.9512 \\
		LINE-1st 			& 0.8091 						& 0.8250 			& 0.4872 					& 0.6639 						& 0.9494 				& 0.9569   \\
		LINE-2nd 			& 0.7559 					& 0.7500  				& 0.5304 					& 0.7377  						& 0.9468  			& 0.9491 \\
		PTE 					& 0.8852 				& 0.8750  					& 0.5389 					& 0.7342  					& 0.9791 				 & 0.9847 \\
		ESim 					& 0.8867 				& 0.8750 					& 0.6836 					& 0.7399 				& 0.9910 				& 0.9948 \\
		HIN2Vec 			& 0.8631 				& 0.8500 					& 0.6075 					& 0.7361 					& \textbf{0.9962} 		& \textbf{0.9965} \\
		metapath2vec & 0.8976 					& 0.9000 					& 0.5337 					& 0.7208 					& 0.9934 			& 0.9936 \\
		
		RHINE 				& \textbf{0.9344} 		& \textbf{0.9250} 		& \textbf{0.7132} 		& \textbf{0.7572} 		& 0.9884		& 0.9807 \\
		\bottomrule
	\end{tabular}
\end{table*}
\subsubsection{Experimental Settings}
We leverage K-means to cluster the nodes and evaluate the results in terms of normalized mutual information (NMI) \cite{shi2014hetesim}. 
\subsubsection{Results}
As shown in Table \ref{tab:node-clustering}, our model RHINE significantly outperforms all the compared methods. (1) Compared with the best competitors, the clustering performance of our model RHINE improves by 18.79\%, 6.15\% and 7.84\% on DBLP, Yelp and AMiner, respectively. It demonstrates the effectiveness of our model RHINE by distinguishing the various relations with different structural characteristics in an HIN. In addition, it also validates that we utilize appropriate models for different categories of relations.
(2) In all baseline methods, homogeneous network embedding models achieve the lowest performance, because they ignore the heterogeneity of relations and nodes. (3) RHINE significantly outperforms existing HIN embedding models (i.e., ESim, HIN2Vec and metapath2vec) on all datasets. We believe the reason is that our proposed RHINE with appropriate models for different categories of relations can better capture the structural and semantic information of HINs.

\subsection{Link Prediction}
\subsubsection{Experimental Setting}
We model the link prediction problem as a binary classification problem that aims to predict whether a link exists. In this task, we conduct co-author (\textit{A-A}) and author-conference (\textit{A-C}) link prediction for DBLP and AMiner. For Yelp, we predict user-business (\textit{U-B}) links which indicate whether a user reviews a business. We first randomly separate the original network into training network and testing network, where the training network contains 80\% relations to be predicted (i.e., \textit{A-A}, \textit{A-C} and \textit{U-B}) and the testing network contains the rest. Then, we train the embedding vectors on the training network and evaluate the prediction performance on the testing network. 
\subsubsection{Results}
The results of link prediction task are reported in Table \ref{tab:lp} with respect to AUC and F1 score. It is clear that our model performs better than all baseline methods on three datasets. The reason behind the improvement is that our model based on Euclidean distance modeling relations can capture both the first-order and second-order proximities. In addition, our model RHINE distinguishes multiple types of relations into two categories in terms of their structural characteristics, and thus can learn better embeddings of nodes, which are beneficial for predicting complex relationships between two nodes. 

\subsection{Multi-Class Classification}
\subsubsection{Experimental Setting}
In this task, we employ the same labeled data used in the node clustering task. After learning the node vectors, we train a logistic classifier with 80\% of the labeled nodes and test with the remaining data. We use Micro-F1 and Macro-F1 score as the metrics for evaluation \cite{dong2017metapath2vec}.
\subsubsection{Results}
We summarize the results of classification in Table \ref{tab:classification}. As we can observe, (1) RHINE achieves better performance than all baseline methods on all datasets except Aminer. It improves the performance of node classification by about 4\% on both DBLP and Yelp averagely. In terms of AMiner, the RHINE performs slightly worse than ESim, HIN2vec and metapath2vec. This may be caused by over-capturing the information of relations \textit{PR} and \textit{APR} (\textit{R} represents references). Since an author may write a paper referring to various fields, these relations may introduce some noise. (2) Although ESim and HIN2Vec can model multiple types of relations in HINs, they fail to perform well in most cases. Our model RHINE achieves good performance due to the respect of distinct characteristics of various relations. 
\begin{figure*}
	\centering
	\begin{minipage}[t]{0.3\textwidth}
		\centering
		\includegraphics[width=1\columnwidth]{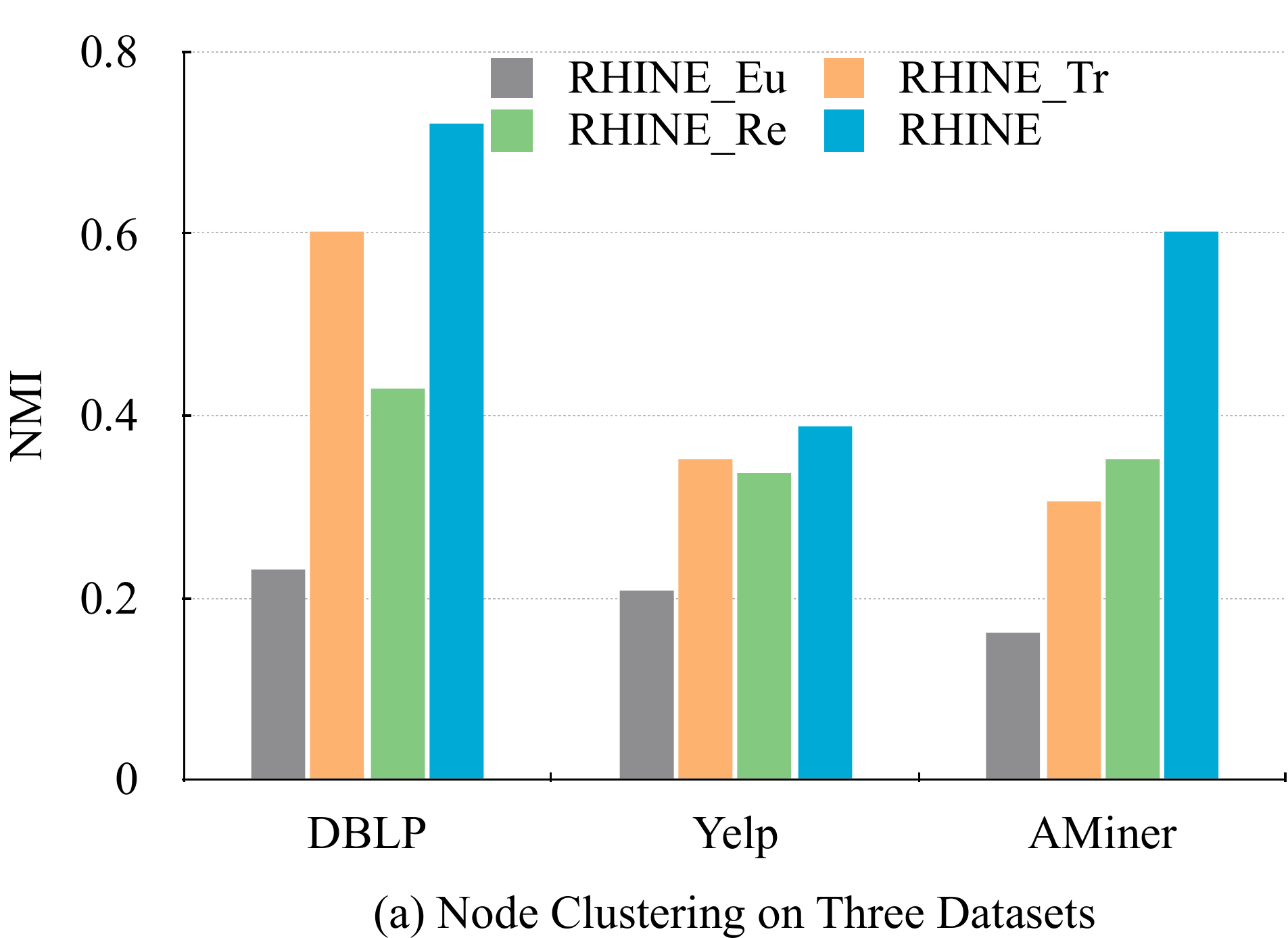}
	\end{minipage}
	\begin{minipage}[t]{0.3\textwidth}
		\centering
		\includegraphics[width=1\columnwidth]{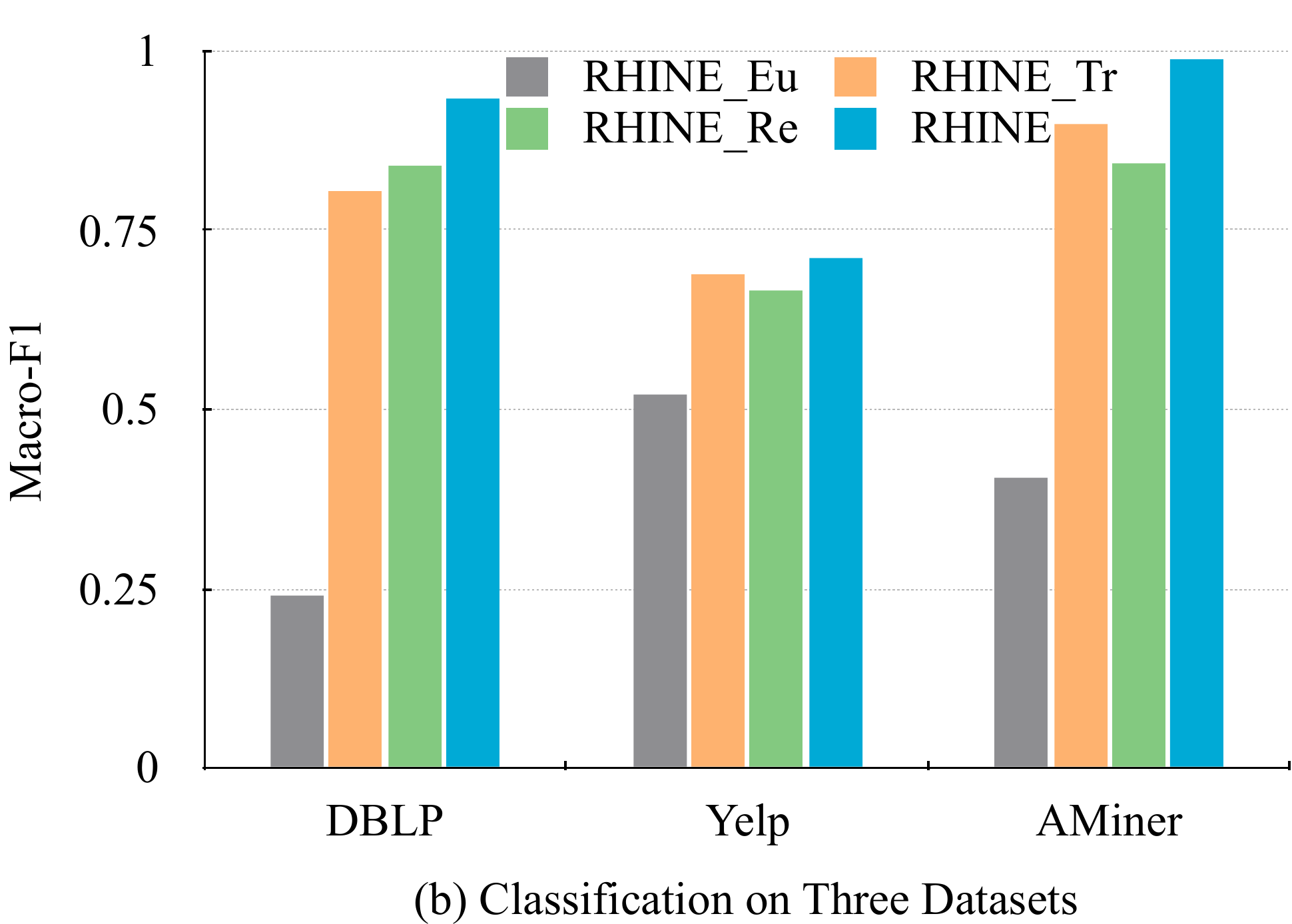}
	\end{minipage}
	\begin{minipage}[t]{0.3\textwidth}
		\centering
		\includegraphics[width=1\columnwidth]{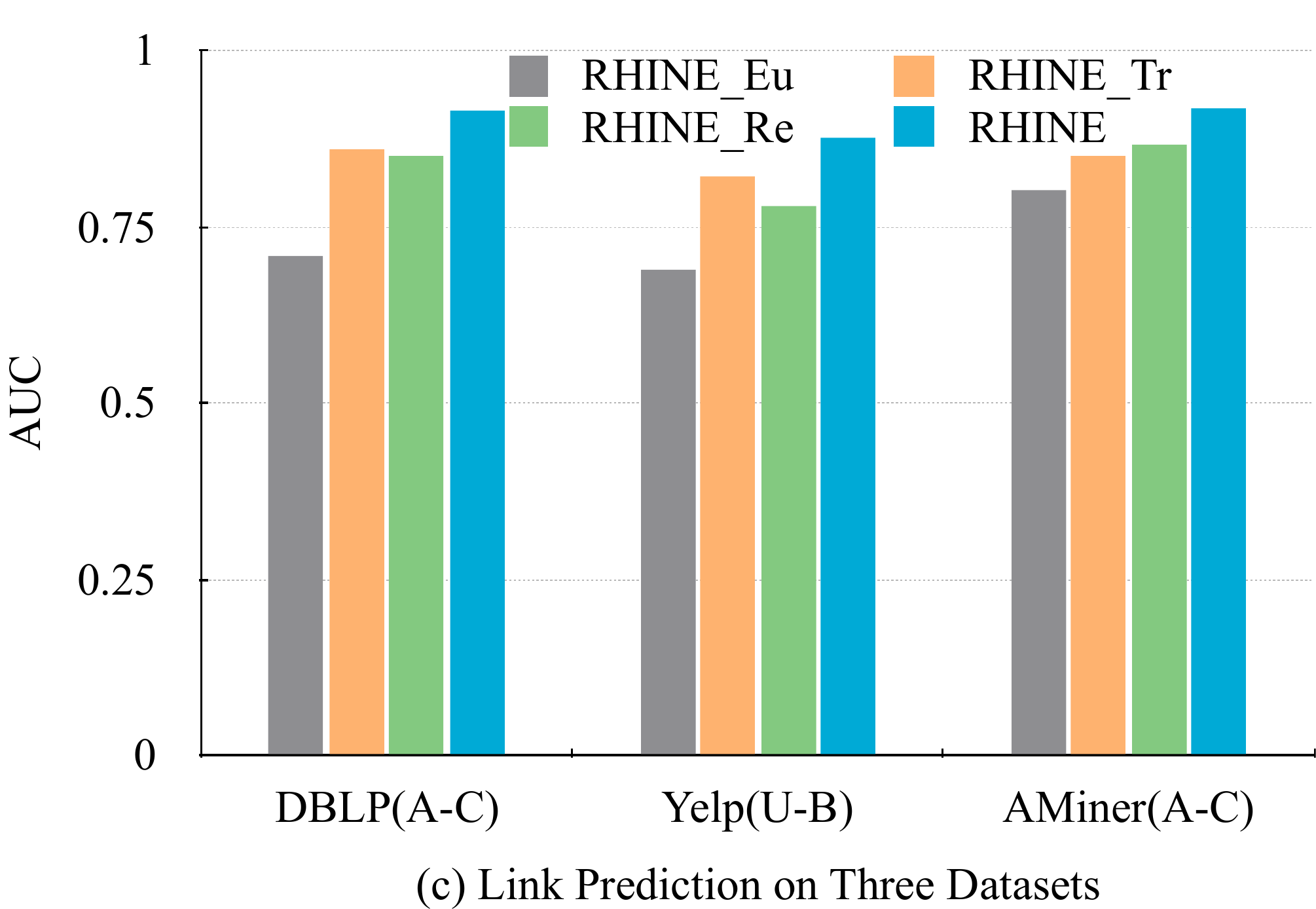}
	\end{minipage}
	\caption{Performance Evaluation of Variant Models.}
	\label{fig:seven}
\end{figure*}
\begin{figure*}
	\centering
	\includegraphics[width=0.92\linewidth]{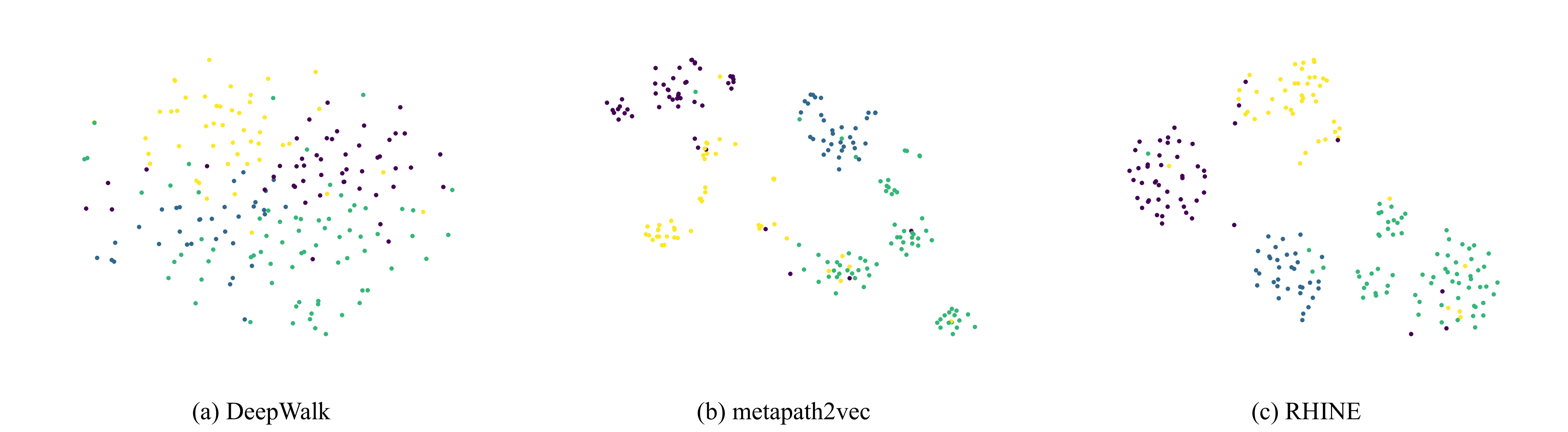}
	\caption{Visualization of Node Embeddings.}
	\label{fig:six}
\end{figure*}

\subsection{Comparison of Variant Models }
In order to verify the effectiveness of distinguishing the structural characteristics of relations, we design three variant models based on RHINE as follows:
\begin{itemize}
	\item 
	$ \textbf{RHINE}_{Eu} $ leverages Euclidean distance to embed HINs without distinguishing the relations.
	\item 
	$ \textbf{RHINE}_{Tr} $ models all nodes and relations in HINs with translation mechanism, which is just like TransE \cite{bordes2013translating}.
	\item 
	$ \textbf{RHINE}_{Re} $ leverages Euclidean distance to model IRs while translation mechanism for ARs, reversely.
\end{itemize}
We set the parameters of variant models as the same as those of our proposed model RHINE. The results of the three tasks are shown in Figure~\ref{fig:seven}. It is evident that our model outperforms RHINE$_{Eu} $ and RHINE$_{Tr} $, indicating that it is beneficial for learning the representations of nodes by distinguishing the heterogeneous relations.
Besides, we find that RHINE$ _{Tr} $ achieves better performance than RHINE$ _{Eu} $. This is due to the fact that there are generally more peer-to-peer relationships (i.e., IRs) in the networks.
Directly making all nodes close to each other leads to much loss of information.
Compared with the reverse model RHINE$_{Re} $, RHINE also achieves better performance on all tasks, which implies that two models for ARs and IRs are well designed to capture their distinctive characteristics.

\subsection{Visualization}
To understand the representations of the networks intuitively, we visualize the vectors of nodes (i.e., papers) in DBLP learned with DeepWalk, metapath2vec and RHINE in Figure~\ref{fig:six}. 
As we can see, our model clearly clusters the paper nodes into four groups. It demonstrates that our model learns superior node embeddings by distinguishing the heterogeneous relations in HINs. In contrast, DeepWalk barely splits papers into different groups. Metapath2vec performs better than DeepWalk, but the boundary is blurry.
\begin{figure}[!tbp]
	\centering
	\begin{minipage}[c]{0.49\columnwidth}
		\centering
		\includegraphics[width=1\linewidth]{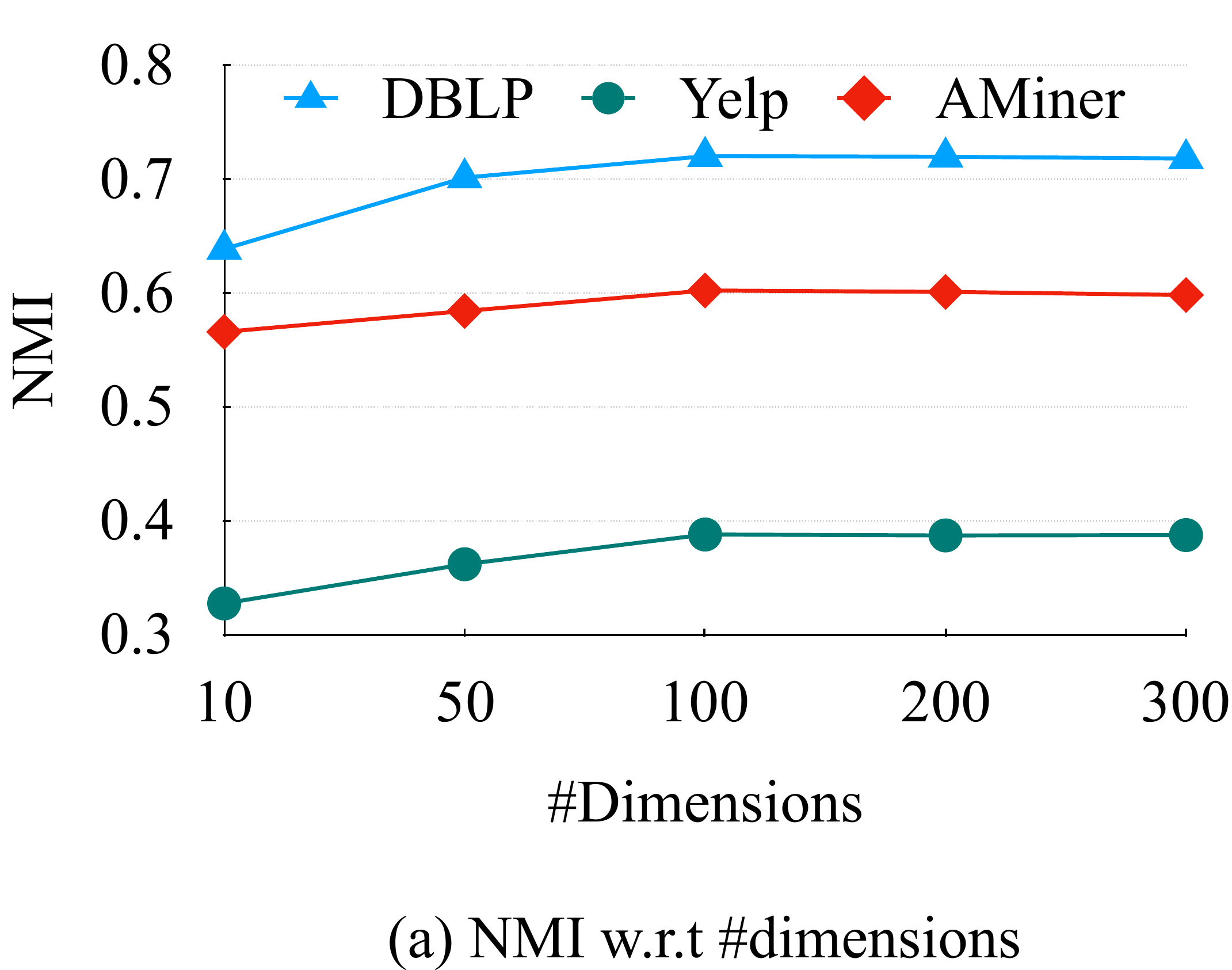}
	\end{minipage}
	\begin{minipage}[c]{0.49\columnwidth}
		\centering
		\includegraphics[width=1\linewidth]{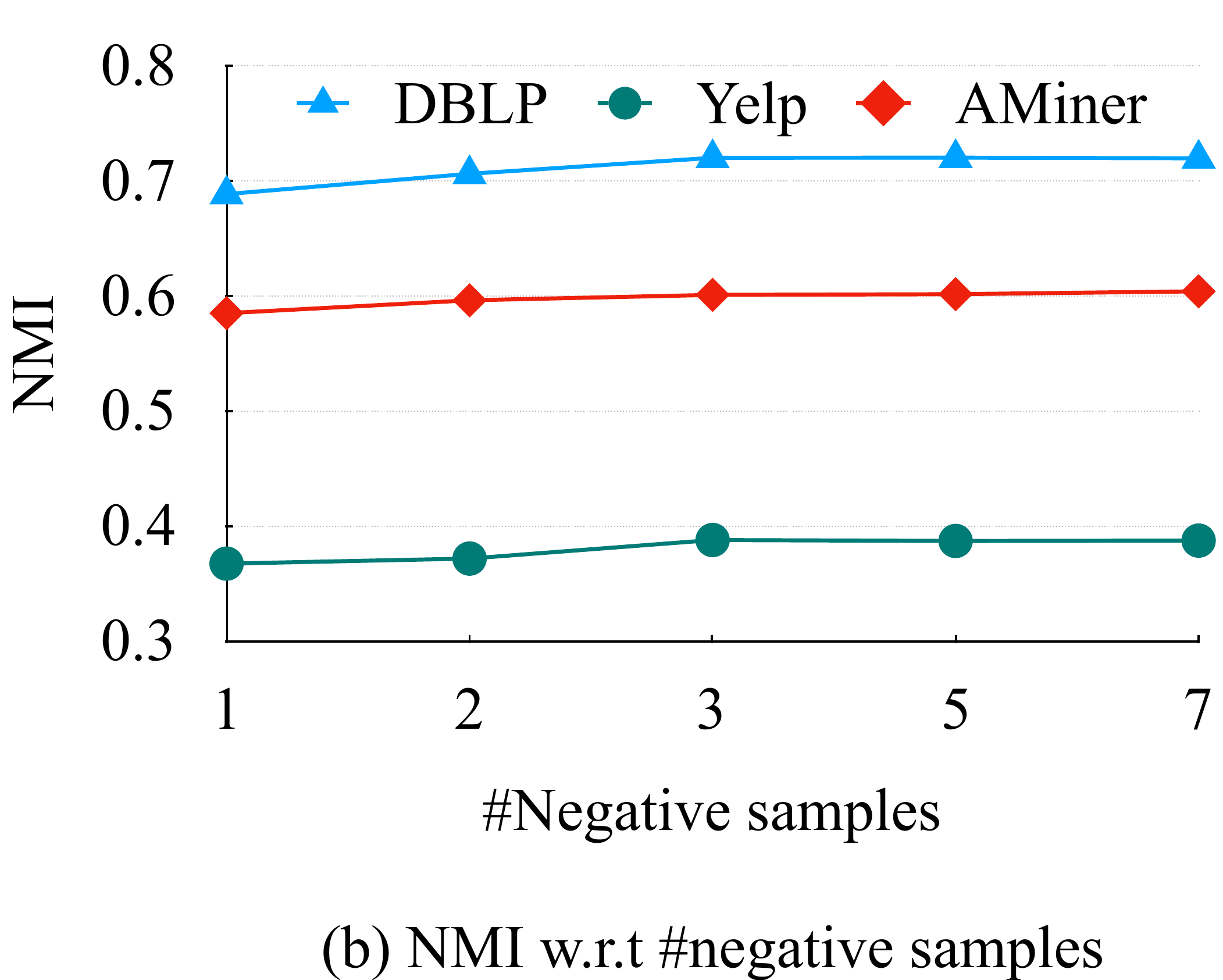}
	\end{minipage}
	\caption{Parameter Analysis.}
	\label{fig:para-tun}
\end{figure}
\subsection{Parameter Analysis}
In order to investigate the influences of different parameters in our model, we evaluate the RHINE in node clustering task. Specifically, we explore the sensitivity of two parameters, including the number of embedding dimensions  and the number of negative samples. 
As shown in Figure~\ref{fig:para-tun}(a), the performance of our model improves with the increase in the number of dimensions, and then tends to be stable once the dimension of the representation reaches around 100.
Similarly, Figure~\ref{fig:para-tun}(b) shows that as the number of negative examples increases, the performance of our model first grows and then becomes stable when the number reaches 3.

\section{Conclusion}
In this paper, we make the first attempt to explore and distinguish the structural characteristics of relations for HIN embedding. We present two structure-related measures which can consistently distinguish heterogeneous relations into two categories: affiliation relations and interaction relations.
To respect the distinctive structures of relations, we propose a novel relation structure-aware HIN embedding model (RHINE), which individually handles these two categories of relations. Experimental results demonstrate that RHINE outperforms state-of-the-art baselines in various tasks.
In the future, we will explore other possible measures to differentiate relations so that we can better capture the structural information of HINs. In addition, we will exploit deep neural network based models for different relations.

\section{Acknowledgments}
This work is supported by the National Key Research and Development Program of China (2017YFB0803304), the National Natural Science Foundation of China (No. 61772082, 61806020, 61702296, 61375058), the Beijing Municipal Natural Science Foundation (4182043), and the CCF-Tencent Open Fund.

\fontsize{9.5pt}{10.5pt} 
\selectfont
\bibliographystyle{aaai} 
\bibliography{AAAI-LuY.5171}

\end{document}